\documentclass[12pt]{article}
\usepackage{latexsym}
\usepackage{epsf}

\thicklines                         
\def\a{\alpha}
\def\b{\beta}
\def\c{\chi}
\def\d{\delta}
\def\e{\epsilon}                
\def\g{\gamma}
\def\h{\eta}

\def\j{\psi}
\def\k{\kappa}
\def\l{\lambda}
\def\m{\mu}
\def\n{\nu}

\def\p{\pi}                     
\def\th{\theta}                  
\def\r{\rho}                    

\def\D{\Delta}

\def\L{\Lambda}

\def\S{\Sigma}

\def\ci{{\cal I}}

\def\cp{{\cal P}}

\def\cbo{{\,\raise-.15ex\Sc [\,}}                       
\def\ltap{\raisebox{-.4ex}{\rlap{$\sim$}} \raisebox{.4ex}{$<$}}   
\def\gtap{\raisebox{-.4ex}{\rlap{$\sim$}} \raisebox{.4ex}{$>$}}   

\def\sl#1{\rlap{\hbox{$\mskip 1 mu /$}}#1}      

\def\sbra#1{\left\langle #1\right|}             
\def\sket#1{\left| #1\right\rangle}             
\def\svev#1{\left\langle #1\right\rangle}       

\def\ddt#1{{\buildrel {\hbox{\LARGE .\kern-2pt.}} \over {#1}}}

\def\beq{\begin{equation}}
\def\eeq{\end{equation}}
\def\bqry{\begin{eqnarray}}
\def\eqry{\end{eqnarray}}

\def\secteq#1{ \setcounter{equation}{0}
               \renewcommand{\theequation}{#1.\arabic{equation}} }

\def\beqn#1{ \renewcommand{\theequation}{#1}
             \begin{eqnarray} }
\def\eeqn{ \renewcommand{\theequation}{\arabic{equation}}
           \end{eqnarray} }

\def\beqr#1{ \setcounter{equation}{#1}
             \begin{eqnarray} }

\def\eeqr{\end{eqnarray}}
\def\NON{\nonumber\\}
\def\beqrabc#1{ \setcounter{equation}{0}
                \renewcommand{\theequation}{#1\alph{equation}}
                \begin{eqnarray} }
\def\beqrn#1#2{ \setcounter{equation}{#2}
                \renewcommand{\theequation}{#1.\arabic{equation}}
                \begin{eqnarray} }
\def\seeq#1{eq.~(\ref{#1})}

\def\seeqs#1{eqs.~(\ref{#1})}

\def\seneq#1{~(\ref{#1})}

\def\NPB#1{Nucl. Phys. {\bf B#1}}
\def\NPBP#1{Nucl. Phys. (Proc. Suppl.) {\bf B#1}}
\def\PLB#1{Phys. Lett. {\bf B#1}}
\def\PRD#1{Phys. Rev. {\bf D#1}}

\def\PRL#1{Phys. Rev. Lett. {\bf #1}}

\def\sstyle{\scriptstyle}

\def\rhs{\mbox{r.h.s.} }

\def\eg{\mbox{e.g.} }
\def\etc{\mbox{etc.} }

\def\frac#1#2{ {\sstyle {#1\over #2} } }

\def\tr{{\rm tr}\,}

\def\half{{1\over 2}}

\def\rcite#1{ref.~\cite{#1}}

\def\qeff{q_{\rm eff}}
\def\ceff{\c_{\rm eff}}
\def\qpt{q_{\rm pt}}
\def\cpt{\c_{\rm pt}}
\def\ns{{N_s}}
\def\GeV{\;{\rm GeV}}

\begin{document}
\hyphenation{fer-mio-nic per-tur-ba-tive}

March 2000 \hfill TAUP--2626--00

\begin{center}
\vspace{15mm}
{\large\bf New Domain-Wall Fermion Actions}
\\[15mm]
Yigal Shamir
\\[5mm]
{\it School of Physics and Astronomy\\
Beverly and Raymond Sackler Faculty of Exact Sciences\\
Tel-Aviv University, Ramat~Aviv,~69978~ISRAEL}\\
shamir@post.tau.ac.il
\\[15mm]
{ABSTRACT}
\\[2mm]
\end{center}

\begin{quotation}
In perturbation theory, the wave function of domain-wall quarks
decreases exponentially with the fifth coordinate.
We show that, regardless of the quark's own momentum,
the fall-off rate of the one-loop
wave function is equal to the {\it slowest} rate encountered
at tree-level for any lattice four-momentum.
We propose new domain-wall actions involving
beyond-nearest neighbor couplings in the four physical dimensions,
for which the perturbative wave function decreases much faster.
It is hoped that the new actions may preserve the
good chiral properties of domain-wall fermions
up to larger values of the lattice spacing.
\end{quotation}



\newpage
\noindent {\large\bf 1.~~Introduction}
\vspace{3ex}
\secteq{1}

The coupling of the right-handed and the left-handed components of
Wilson fermions through the QCD interaction leads
in the continuum limit to the chiral anomaly~\cite{ks},
but for finite lattice spacing $a$ it also leads to
lattice-artefact violations of chiral symmetries.
This results in an additive renormalization of the quark mass,
as well as in a severe tuning problem for four-fermion operators
which are needed for the computation of weak matrix elements.
The mass renormalization is $O(g^2_0)$ in lattice units.
Since $g_0^2 \sim (\log(a\, \L_{\rm QCD}))^{-1}$,
the mass renormalization diverges like
$(a \log(a\, \L_{\rm QCD}))^{-1}$ in the continuum limit $a \to 0$.
In the challenging lattice calculation of non-leptonic kaon decays
(e.g.\ $\e'/\e$)
the tuning problem is formidable because of the large number of
chirality-disallowed mixings.

In the domain-wall formulation of lattice QCD~[2-6], the two
chiral components arise as surface
states on opposite boundaries of a five-dimensional lattice,
and one expects their coupling to vanish
when the size of the fifth dimension tends to infinity.
(The fifth coordinate will be denoted $s$, and it
takes values $0 \le s \le \ns-1$.)
A precise non-perturbative characterization of chiral symmetry violations
can be given in terms of the transfer matrix for hopping in the
$s$-direction~\cite{nn1,fs}.
Being the result of slow decay of correlations in the $s$-direction,
chiral symmetry violations are associated with near-unity eigenvalues
of that transfer matrix.
For a given gauge-field configuration the approach to the chiral limit
is exponential {\it iff} the spectrum of the transfer matrix has a gap,
and the fall-off rate is determined by the size of the gap.

In full QCD
there are several analytic results concerning the $\ns \to \infty$ limit.
The approach to the chiral limit is exponential
in perturbation theory~\cite{sh,at,kny},
and the same is true non-perturbatively if a constrained gauge action
(believed to be in the same universality class as the standard
plaquette action) is used~\cite{constr}.

For an unconstrained action one can also prove non-perturbatively
that chiral symmetry is restored in the limit $\ns \to \infty$,
provided the (finite!) number of sites in each
of the four physical lattice dimensions is held fixed~\cite{fs}.
When near-unity eigenvalues start playing a significant
role, the chiral limit may be approached as slow as $1/\ns$.
The proof that certain symmetries are restored in the  limit $\ns \to \infty$
is actually valid  for {\it any} value of the coupling constant.
But the identification of the restored symmetries as chiral ones depends
on the fermion spectrum. It was recently shown~\cite{rbbs}
that within the strong-coupling expansion the massless spectrum
of the domain-wall lattice hamiltonian is either doubled or empty.
Therefore the restored symmetries are not chiral at strong coupling
(For further details see Appendix~C.1).

Of major importance is the question of how close to the chiral limit
one gets in Monte-Carlo simulations.
In trying to answer this question we rely on two sources.
The first is the spectrum of the transfer matrix,
or of the closely-related~\cite{nn1,redch} hermitian Wilson-Dirac operator.
In the latter case, a key finding~\cite{scri} is that the
{\it spectral density} of near-zero modes
(corresponding to near-unity eigenvalues of the transfer matrix)
rises by two orders of magnitude as the (quenched) coupling
changes from $6/g^2\equiv\b=6.3$ ($a^{-1} \sim 4 \GeV$) to $\b=5.7$
($a^{-1} \sim 1 \GeV$).
We hope that more results on the eigenvalue spectrum
will be available in the future.

More information is available through lattice computations
of various correlation functions~[13-20].
A detailed numerical study of the chiral limit of
domain-wall fermions was first carried out
in the Schwinger model~\cite{pv}.
In QCD the first domain-wall simulations were promising,
and the results for weak matrix elements ($B_K$, $O_{LL}$) \cite{bs}
and for the strange-quark mass~\cite{bsw}
were in agreement with other methods.
As of today more data is available.
The pion-mass squared, which should extrapolate linearly
to zero with the quark mass, is
the most obvious measure of chiral symmetry. Using domain-wall fermions
at quenched $\b=6.0$ ($a^{-1} \sim 2 \GeV$) the extrapolated pion mass
(for $\ns\sim 20$) does not vanish exactly at zero quark mass but, rather,
at a negative value of the order of few times $10^{-3}$
in lattice units~\cite{priv} (see also \rcite{cppacs}).
This value, however small,
is in the same range as the light quark masses.

In a sense, chiral symmetry violations are worst for the pion mass,
because the lattice-artefact term in the PCAC relation
is an ensemble average of a positive fermion correlator
(see Appendix~C.1; we comment that
that lattice-artefact term is a better measure of chiral symmetry violation
compared to the extrapolated pion mass, since it does not suffers from
theoretical uncertainties due
to chiral perturbation theory and due to finite-volume effects).

No such positivity is encountered in the calculation of
weak matrix elements~\cite{bs,chd},
so chiral symmetry violations are expected to be smaller in this case.
For example,
in a recent simulation of four-fermion operators using the non-perturbative
renormalization scheme, again at quenched $\b=6.0$, and using $\ns=16$,
it was found~\cite{chd} that mixing into wrong-chirality operators
was practically zero (in comparison with 10\% for Wilson fermions
in a typical example).

Going to a smaller value of the inverse lattice spacing,
the situation at $a^{-1} \sim 1 \GeV$ is unsatisfactory,
as deviations from chiral symmetry are significant even for $\ns$ as large as
50 or 100 (in both quenched and dynamical simulations) \cite{57i,57}.
In the opposite direction, at  $a^{-1}\; \gtap \; 3 \GeV$,
no difficulties with the restoration of chiral symmetry have been
reported~\cite{bs,scri,constr}.

We believe that the existing results, especially those
for weak matrix elements at $a^{-1} \sim 2 \GeV$,
do represent a breakthrough compared to the ``pre domain-wall era''.
On the other hand, the results for the pion mass at $a^{-1} \sim 2 \GeV$ are
not as good as one would hope for, and the present situation
at $a^{-1} \sim 1 \GeV$
makes scaling studies with domain-wall fermions very difficult.

Having summarized the situation in Monte-Carlo simulations
let us return to the underlying physics.
The key question is what mechanism(s) determine
the abundance of near-unity eigenvalues of the transfer matrix.
It is known that a few exact-unity eigenvalues must occur
during the transition from one topological sector to another
on a finite lattice with periodic boundary conditions \cite{nn1,nn2}.
We believe, however, that the role of topology changing has been
over-emphasized, for topological considerations alone
do not explain the proliferation of near-unity eigenvalues
nor the magnitude of the ensuing chiral symmetry violations.

A simple explanation may be that the observed
chiral symmetry violations arise (mainly) from generic fluctuations of
the gauge field~\cite{tom}.
The effect of fluctuations need not be small,
because the coupling constant used in simulations is not small either.
At the same time, as long as the coupling constant has not grown too much,
perturbation theory should provide a reliable approximation of
the leading quantum effects.

In this paper we calculate the fifth coordinate's wave function
of domain-wall quarks in the one-loop approximation (Sec.~2).
The results lead us to consider new classes of domain-wall actions (Sec.~3).
A preliminary account of this work was given in \rcite{dubna}.
(An alternative/complementary approach, whose relative merits
are discussed in Sec.~4, is to employ an improved
gauge action~\cite{57i}.)

We now give an overview of the one-loop calculation.
For large $\ns$, zero bare
quark mass and with the right-handed quark field near the $s=0$
boundary, the dressed fermion propagator near that boundary is
\beq
  G_{s,s'}(p) = P_+\; \c(s)\, {1\over i\sl{p}(1+\S_K)}\, \c(s')
  + {\rm Reg.,}\qquad s,s' \ll \ns \,,
\label{G}
\eeq
where $P_{\pm}={1\over 2}(1\pm \g_5)$,
``Reg'' stands for a continuous function of the four-momentum
$p$, and $\S_K \approx \S_K(g^2,g^2 \log(p^2))$.
A unique feature of the domain-wall scheme is $\c(s)$, the $s$-coordinate
wave function for (right-handed) quark modes.
At tree level one has
\beq
  \c_0(s) \propto q_0^s \equiv (1-M)^s \,.
\label{c0}
\eeq
(The five-dimensional mass term $M$  is often
referred to as the domain-wall height, and should not
be confused with the quark mass~\cite{sh}.) By choosing $M=1$ the free
wave function can be completely localized on the boundary
\beq
  \c_0(s) = \lim_{M\to 1}(1-M)^s = \d_{s,0} \,.
\label{c0M1}
\eeq
The result of the one-loop calculation of the wave function is
\beq
  \c_1(s) \sim s^{-2}\, q_1^s \,,
\label{c1}
\eeq
which contains also a power correction.
Like an ordinary Wilson mass, $M$ is renormalized additively.
Making an optimal choice of $M$ we find
\beq
 q_1 = 0.5 \,.
\label{qpt}
\eeq
It should be noted that the difference between $q_0$ and $q_1$ is $O(1)$.
In the full one-loop result (\seeqs{wf1} to\seneq{result} below)
$g^2$ occurs as a pre-factor
of relatively little importance.

Let us now explain the physical origin of $q_1$.
Consider the free domain-wall propagator $G^0_{s,s'}(p)$
for a given four-momentum $p$ in the vicinity of the $s=0$
boundary at zero quark mass. The $s$-correlations described by
this propagator are controlled by an exponent $\a(p)$.
Each term in the propagator involves a factor $\exp(-d\,\a(p))$
where $d$ stands for either
the separation $|s-s'|$ or the sum of distances from the boundary
$s+s'$. For the standard domain-wall action one has
${\rm max} \{\exp(-\a)\} = 0.5$ for $M=1$ where the maximum
over the Brillouin zone is obtained at the ``corner'' $p_\p=(\p,0,0,0)$
and its three permutations.
We will denote the set of global maxima by $\cp$.

Now, at tree level, a fermion eigenmode with momentum $p$ propagates
independently of all other eigenmodes.
But for any non-zero gauge coupling
the fermions propagate in non-trivial backgrounds,
and these backgrounds allow any given momentum eigenmode to
couple to all momentum eigenmodes.
In particular, small-momentum quark modes couple to modes with
$p \in \cp$.

We arrive at the following physical picture.
A four-dimensional fermion mode created on a given $s$-layer
mixes on that layer with the modes of $\cp$ through the gauge field.
As a mode with $p \in \cp$,
the fermion propagates with minimal suppression to
some other layer $s'$, where the action of the gauge field
turns it back into the original mode. Propagation
in the $s$-direction is therefore dominated by the
modes of $\cp$ leading to
\beq
  q_1 = {\rm max} \{\exp(-\a(p))\} \,.
\label{q1}
\eeq
For $M=1$ this reduces to \seeq{qpt}.
If we would momentarily regard
the fifth direction as an imaginary-time direction,
the above is recognized as the familiar result
that propagation is always dominated by the lightest excitation
in any given channel. The domain-wall case is particularly simple
in that the gauge field is independent of the $s$-coordinate.

Under certain conditions (basically that the coupling constant
is not too large, see Appendix~C.2 for a more detailed discussion) it should be
possible to describe the results of numerical simulations too
in terms of an effective wave function $\ceff(s) \sim s^{-1-\d} \qeff^s$.
This means that every quark's wave function is assumed to be the product of
a four-dimensional wave function and the universal fifth-coordinate
wave function $\ceff(s)$. The exponential fall-off rate is
accounted for by $\qeff$.
At relatively weak coupling ($a^{-1} \;\gtap\; 3 \GeV$) there seems to
be no problem with the restoration of chiral symmetry,
suggesting that $\qeff < 1$.
For $a^{-1} \sim 2 \GeV$, the rate at which chiral symmetry is restored
depends sensitively on the observable. This, as well as other indications,
suggest that $\qeff$ is very close to one, and the restoration
of chiral symmetry really follows a power-law behavior.
(In \rcite{dubna} an estimate of $\qeff$ was given
which, however, is unjustified because the power-law correction was ignored.)
Then, at $a^{-1} \sim 1 \GeV$ the notion of a universal, localized,
$s$-coordinate wave function breaks down.

Comparing the perturbative results with the numerical data
shows that, not surprisingly, the optimal tree-level value $q_0=0$
completely fails to describe that data.
In comparison, the one-loop result $q_1=0.5$ lies approximately ``half-way''
between the tree-level value and the close-to-one values
of $\qeff$ which seem to account for the results of simulations.
Thus $q_1$ gives at least some indication of the actual
behavior of the system.

In this paper we adopt $q_1$ as an analytic criterion for
the quality of domain-wall actions.
In Sec.~3 we consider new families of domain-wall actions
involving beyond-nearest neighbor coupling.
We compute the resulting $q_1$, and find that values much smaller than
0.5 can be achieved.
Finally, in Sec.~4 we discuss the relevance of our results
to numerical simulations.

Some technicalities of the one-loop calculation are relegated
to Appendix~A. Higher order corrections are briefly discussed in Appendix~B.
An expanded discussion of some non-perturbative issues
can be found in Appendix~C. 

\vspace{5ex}
\noindent {\large\bf 2.~~The one-loop wave function}
\vspace{3ex}
\secteq{2}

In this section we calculate the one-loop wave function of
domain-wall quarks, relegating some of the technicalities to Appendix~A.
The finite-$\ns$ tree-level propagator and the one-loop
self energy for domain-wall fermions were calculated in \rcite{pv,at}
(see also \rcite{kny}).
Here we will be interested in the range $1 \ll s \ll \ns$,
therefore we can use the simpler expressions for the tree-level propagator
in the limit $\ns\to\infty$ \cite{sh}.
Assuming the right-handed quark is localized near the
$s=0$ boundary, the singular part of the tree-level propagator
$G^0_{s,t}(p)$ is
\beq
   P_+ \; {M(2-M)(1-M)^{s+t}\over i\sl{p}} \,.
\label{G0M}
\eeq
For $M\to 1$ this becomes
\beq
  P_+ \; {\d_{s,0}\,\d_{t,0}\over i\sl{p}} \,,
\label{G0}
\eeq
which means that the massless right-handed fermion field
is fully localized on the boundary layer.

The first quantum effect beyond the free theory
is an additive correction, $-\d M$, to the five-dimensional mass $M$.
It arises in a mean-field approximation, or in perturbation theory
from tadpole diagrams. It is well known that this effect
must be treated non-perturbatively~\cite{lm}.
We thus discard the tadpole diagrams, absorbing them into
the tree-level action via the replacement $M \to M-\d M$.

The full one-loop wave function is
\beq
  \c_1(s) = |1+\d M - M|^s +  \d\c_1(s) \,,
\label{wf1}
\eeq
where $\d\c_1(s)$ comes from the ``setting sun'' diagram only.
We will find $\d\c_1(s)$ by matching the non-analytic piece
of the dressed propagators with the \rhs of \seeq{G}.
We are interested in the behavior of $\d\c_1(s)$ when $M$ is
close to its optimal mean-field value (see also Sec.~3).
In the calculation below we thus set  $M=1+\d M$
in tadpole-improved perturbation theory.
Since the tadpole-improved free propagator is a function
of $M-\d M$ (and not of $M$ and $\d M$ separately)
the resulting propagator is identical to the ordinary
propagator with $M=1$.
The setting-sun diagram will therefore be computed
using the expression for the ordinary tree-level propagator for $M=1$.

\begin{figure}[b]
\vspace*{0.8cm}
\centerline{
\epsfxsize=6.0cm
\epsfbox{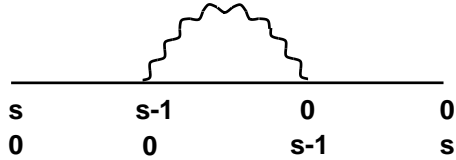}
}
\vspace*{0.0cm}
\caption{ \noindent {\it The ``setting-sun'' diagram.
The external lines are not amputated.
The fifth coordinate of each point, as indicated on the first row below
the diagram, corresponds to the case where $1/\sl{p}$ (\seeq{G0})
occurs on the rightmost line, and the leftmost point
is far off the boundary. This yields \seeq{c1left}.
The second row corresponds to \seeq{c1right}.
}}
\label{set}
\vspace*{0.5cm}
\end{figure}

The setting-sun diagram is depicted in Fig.~\ref{set}.
Notice that we have not amputated the external legs~\cite{norman}.
Except for $s=t=0$ (see\seneq{G0}), the tree-level propagator is not
singular at $p=0$. To obtain a contribution to the \rhs of \seeq{G},
at least one of the three fermion lines must coincide with
expression\seneq{G0}. (The kinetic self-energy correction $\S_K$ in \seeq{G}
arises when all three lines coincide with \seeq{G0}, see \rcite{at},
and will not be discussed here any further.)
Assume first that the rightmost propagator in Fig.~\ref{set}
coincides with \seeq{G0}.
Since the leftmost coordinate $s$ is by assumption
far from the boundary, we may take the limit $p \to 0$ in the
expressions for the self-energy part and for the leftmost propagator.
We thus arrive at
\beq
  \d\c_1(s) =  g^2\,C_2 \sum_{s'\ge 0} G^{0+}_{s,s'}\, \S^+_{s',0}
  = g^2\,C_2\, \S^+_{s-1,0} \,,
\label{c1left}
\eeq
where $C_2$ is the quadratic Casimir and
\beq
  G^{0\pm}_{s,t} = {1\over 2}\; \tr P_\pm\, G^0_{s,t}(p=0)\,,\qquad
  \S^\pm_{s,t}  = {1\over 2}\; \tr P_\pm\, \S_{s,t}(p=0)\,,
\label{proj}
\eeq
and $\S_{s,t}(p)$ is the 1PI self-energy obtained by amputating
the external legs in Fig.~\ref{set}. In the second
equality of \seeq{c1left} we used the explicit expression for
$G^{0+}_{s,t}$ far from the boundary (see Appendix~A).

When substituting \seeq{wf1} into \seeq{G} we find another term, $\d\c_1(s')$.
This term is obtained when
the leftmost propagator in Fig.~\ref{set} coincides with \seeq{G0}.
Following similar steps we now find
\beq
  \d\c_1(s') = g^2\,C_2\, \S^-_{0,s'-1} \,.
\label{c1right}
\eeq
Thanks to a ``parity'' symmetry (see Appendix~A)
one has $\S^+_{s,t} = \S^-_{t,s}$.
Hence \seeqs{c1left} and\seneq{c1right} agree.

It remains to compute the diagonal part of the self-energy.
One can write
\beq
  \S^+_{s,0} = \int_{-\p}^{+\p} {d^4k\over (2\p)^4}\,
  h^+(k) \exp(-s\,\a(k)) \,.
\label{int}
\eeq
(Note that the external momentum is zero.)
The $s$-dependence enters through the exponential.
All other factors were lumped into $h^+(k)$
(see Appendix~A for more details).
For large $s$, the above integral can be computed using a saddle-point
approximation. As mentioned in the introduction, the global maximum of
$\exp(-\a)$ corresponds to the lattice momentum $p_\p$
and its three permutations. In the computation we take $h^+=h^+(p_\p)$
outside the integral, and expand the exponent to second order around
$p_\p$ where we define $k=(\p+k_\parallel,\vec{k}_\perp)$.
Including a factor of four to account for the degeneracy
of the global maximum we obtain
\bqry
  \S^+_{s,0}
& = & - {3\over 2}\left({1\over 2}\right)^s
      \int {dk_\parallel d^3k_\perp \over (2\p)^4}\,
      \exp\left(-{s\over 24}(7 k_\perp^2 + k_\parallel^2) \right)
\NON
& = & - {54\over \p^2\, 7^{3/2}\, s^2\, 2^s} \,.
\label{half}
\eqry
For the fundamental representation of SU(3) one
has $C_2=4/3$. Substituting in \seeq{c1left} we finally find
\beq
  \d\c_1(s) = - g^2\, {72\over \p^2\, 7^{3/2}\, (s-1)^2\, 2^{s-1}}
\approx - g^2\, {0.788\over s^2}\, \left({1\over 2}\right)^s \,,
\qquad s \gg 1 \,.
\label{result}
\eeq
Extrapolating \seeq{result} to smaller values of $s$ suggests that
as soon as (or shortly after) we move off the boundary layer,
$\d\c_1(s)$ dominates over the tree-level term in \seeq{wf1}.
This is true even if $g^2$ is small
(or had the prefactor in \seeq{result} been numerically small).
The reason is that the relative magnitude
of the two terms is proportional to $(0.5/(1+\d M-M))^s$,
and since $1+\d M-M \ll 1$ this grows exponentially fast.

\vspace{5ex}
\noindent {\large\bf 3.~~New actions}
\vspace{3ex}
\secteq{3}

We have found that the $s$-coordinate's wave function of domain-wall quarks
is dominated by quantum effects.
The arguments of Sec.~2 show that the broadening of the wave-function
is controlled, in the one-loop approximation,
by the maximum of $\exp(-\a(p))$ over the Brillouin zone.
This remains true for other domain-wall actions
unless the $s$-couplings are drastically changed.
So, if a different domain-wall action yields a smaller
${\rm max}\{\exp(-\a)\}$,
namely a faster fall-off of the wave function at the one-loop level,
it is plausible that that new action also performs better
non-perturbatively (we return to this issue in Sec.~4).

The standard domain-wall action contains two parameters,
the domain-wall height $M$ and the Wilson parameter $r$
(which is usually set equal to one). The $M$-dependence of
${\rm max}\{\exp(-\a)\}$ was investigated in \rcite{ktv}.
As mentioned in Sec.~2, however, the additive renormalization
of $M$ must be treated non-perturbatively.
The optimal value used in simulations ($M \sim 1.8$)
is nicely consistent with mean-field estimates.
We will thus assume that the numerical optimization of $M$
corresponds to setting $M=1+\d M$ in tadpole-improved perturbation theory.
Again, this means that we should determine ${\rm max}\{\exp(-\a)\}$
using the tree-level action with $M=1$.
As for the Wilson parameter, changing its value
in the standard domain-wall action turns out to have
little effect (see below).

\setcounter{figure}{0}
\renewcommand{\thefigure}{2\alph{figure}}

\begin{figure}[hbt]
\vspace*{0.4cm}
\centerline{
\epsfxsize=8.0cm
\put(-100,0){\epsfbox{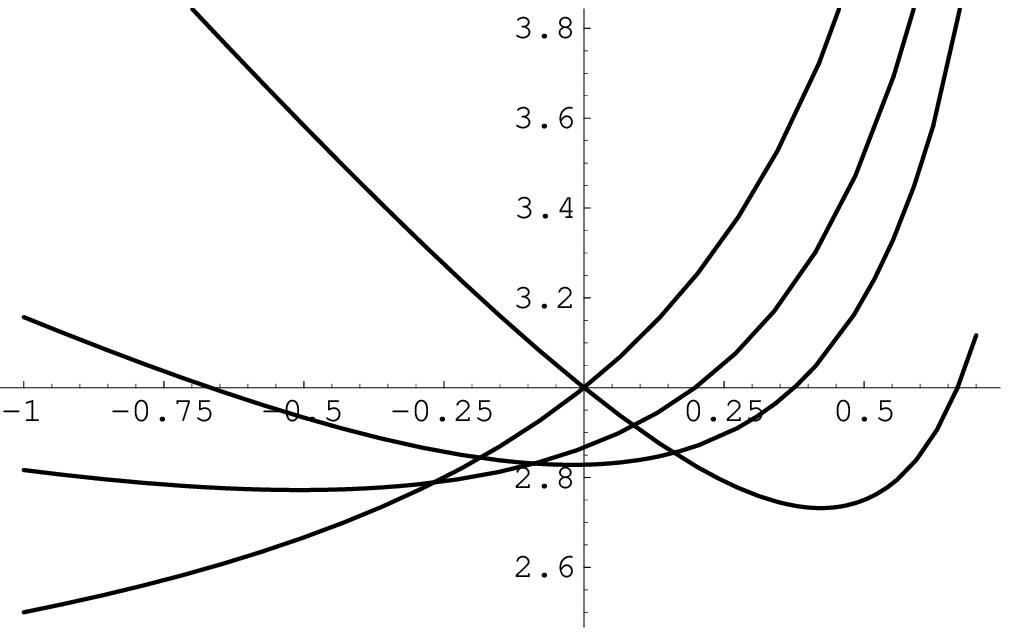}}
\put(65,130){$1.0$}
\put(100,150){$1.2$}
\put(120,115){$1.4$}
\put(120,75){$2.0$}
}
\vspace*{0.4cm}
\caption{ \noindent {\it $2\cosh(\a)$ as a function of
$x=\cos(p_1)$ and $p_2=p_3=p_4=0$ for the standard domain-wall action.
Results are shown for four values of the Wilson parameter $r$.
In all the figures of this section, the minimum
seen in the plot is also the global minimum over the entire
Brillouin zone.
(At larger values of $x$, $2\cosh(\a)$ is monotonically increasing,
and eventually it diverges for $x \to 1$.)
}}
\label{var1}
\vspace*{0.5cm}
\end{figure}

We will depart from the standard domain-wall action by
allowing for couplings not only between nearest neighbors.
In view of the obvious increase in computer time
needed for the inversion of the fermion matrix,
we try to be as economic as possible in our
beyond-nearest neighbor excursion. We allow
only for coupling between sites $x$ and $x+n\hat\m$ (but not
\eg for coupling between $x$ and $x+\hat\m+\hat\n$ for $\m \ne \n$).
In this paper we consider explicitly $n=2$ and $n=3$,
namely next-nearest and next-next-nearest couplings in the same direction.
Also the modifications will be restricted to the four-dimensional part of
the action, leaving the coupling in the fifth direction intact.
(Note that we are interested in achieving a fast fall-off in
the $s$-direction; any attempt to generate a smoother, continuum-like,
behavior in the $s$-direction is thus the exact opposite
of what we are aiming for.)

The domain-wall operators considered here will have the following general
form for zero quark mass
\beq
  D^{\rm d.w.}_{s,t} = \d_{s,t}\, D + (\d_{s+1,t} - \d_{s,t}) P_+
  + (\d_{s-1,t} - \d_{s,t}) P_- \,,
\label{D5}
\eeq
with the understanding that on a finite lattice $0 \le s,t \le \ns-1$.
The inclusion of a quark mass can be done in the usual way~\cite{sh}.
The four-dimensional part of the action is
\beq
  D(p) = i\sum_\m \g_\m\, f(p_\m) - r\, W(p) + M \,.
\label{D}
\eeq
This equation gives the tree-level operator in momentum space.
The generalized Wilson term $W(p)$
is a function of $\cos(p_\m)$. In the kinetic term,
$f(p_\m)$ is an odd function of its argument,
which we take to be $\sin(p_\m)$ times a polynomial in $1-\cos(p_\m)$.
Later we will give explicit expressions for $W(p)$ and $f(p_\m)$.
As explained earlier we set $M=1$ in the tree-level action,
but we will use the freedom in varying the Wilson parameter $r$.
Our convention is that $W(p)$ and $r$ are both positive.

For any domain-wall action of the above form,
the exponents $\a(p)$ are determined by
\beq
  2 \cosh(\a(p)) = {1 + B^2(p)
  + \sum_\m f^2(p_\m) \over B(p)} \,,
\label{cosh}
\eeq
where $B(p)=1-M+r\,W(p)$ and $\a(p)\ge 0$ by convention.
($B(p)=r\,W(p)$ for $M=1$;
for the standard domain-wall action \seeq{cosh} reduces to \seeq{alpha}.)
Lowering the global maximum of $\exp(-\a(p))$
corresponds to raising the global minimum of \seeq{cosh}.

\begin{figure}[thb]
\vspace*{0.4cm}
\centerline{
\epsfxsize=8.0cm
\put(-100,0){\epsfbox{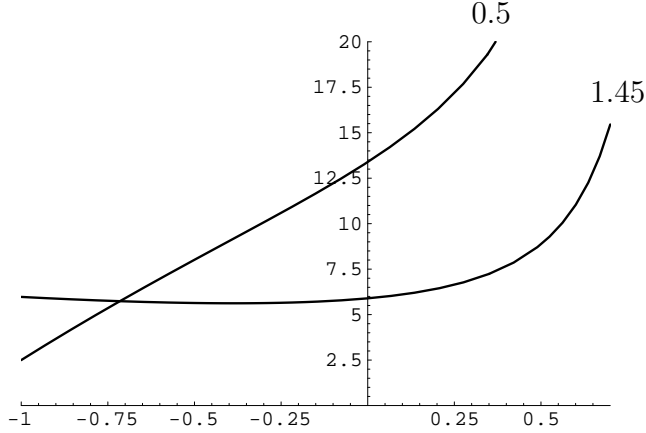}}
\put(75,155){$0.5$}
\put(120,125){$1.45$}
}
\vspace*{0.4cm}
\caption{ \noindent {\it Same as Fig.~\ref{var1} for the action $D_{23}$
with $c_3=4/3$ for two values of $r$. The optimal value is
$r_{\rm opt}=1.45$, whereas $r_2=0.5$ is a ``reference value''
defined by the condition $r_n W_n=2$ at $x=\cos(\p)=-1$ (see Sec.~4).
Notice the difference in vertical scales compared to Fig.~\ref{var1}.
}}
\label{var2}
\vspace*{0.5cm}
\end{figure}

Some insight about the features that control
${\rm min}\{2 \cosh(\a) \}$
can be obtained from very general considerations.
One has $W=0$ for $p=0$, and in all cases one
aims for $r\,W > 1$ at $p_\p=(\p,0,0,0)$.
As we gradually increase $p_1$ for 0 to $\p$ (keeping $p_2=p_3=p_4=0$)
at some value $p_c=(p_{1c},0,0,0)$ we will have $r\,W=1$.
Were it not for the $f(p_1)$ term in \seeq{cosh},
at $p_c$ we would obtain $\cosh(\a) = \exp(-\a) = 1$,
namely no exponential suppression at all.
To avoid this dangerous situation,
we would like to have $f^2(p_1)$ as large as possible at $p_1=p_{1c}$.

Another danger lurks at the (fifteen non-zero)
corners of the Brillouin zone. There, by construction, $f(p_\m)=0$,
and so $\exp(-\a)=(r\,W)^{-1}$. We will therefore also be interested in
increasing $r\,W$ at the corners of the Brillouin zone.

As a warm-up exercise let us consider the effect of varying
$r$ in the standard domain-wall action.
In this case $W=\sum_\m (1- \cos(p_\mu))$ and $f(p_\m)=\sin(p_\m)$.
For $r \le 1$,  ${\rm min}\{2\cosh(\a)\}$
occurs at $p_\p$. We can increase $r\,W$
at $p_\p$ by increasing $r$. But in that case the value
$r\,W=1$ will occur at a smaller $p_1$,
where $\sin(p_1)$ is smaller.
As can be seen from Fig.~\ref{var1} there is a transition region around
$r \sim 1.2$. For larger values of $r$, ${\rm min}\{2\cosh(\a)\}$
moves towards the point where $r\,W=1$.
As an example, for $r=2.0$ one has $r\,W=1$ at $x=\cos(p_1) = 0.5$,
and ${\rm min}\{2\cosh(\a)\}$ is at $x \sim 0.42$.
The largest value of
${\rm min}\{2\cosh(\a)\}$, obtained for $r \sim 1.3$ -- 1.5, is around 2.8.
This makes little improvement over the value 2.5 obtained at $r=1$.

To avoid this dead-lock we take $W$ to be a {\it non-linear} function
of $\cos(p_\mu)$. We define
\beq
  W_n = \sum_\m (1- \cos(p_\mu))^n \,.
\label{Wn}
\eeq
$W_n$ requires couplings of sites $x$ and $x+n\hat\m$.
Once such coupling have been introduced into the generalized
Wilson term, we allow them also in the kinetic term.
Further raising of the global minimum of $\cosh(\a)$ will be made possible
by choosing $f(p_\m)$ that increases faster than $\sin(p_\m)$,
and by adjusting the Wilson parameter $r$.

We now turn to the investigation of concrete actions.
The minimization problem was solved numerically.
Note that the \rhs of \seeq{cosh} can be expressed as a function
of $x_\m \equiv \cos(p_\m)$ only. Using the invariance under
permutations of the four components, it is enough to look for
the global minimum over the range $-1 \le x_1 \le x_2 \le x_3 \le x_4 \le 1$.
(One can also study the minimization problem analytically.
For any $\m$, $\sin(p_\mu)=0$ always satisfies the extremality
condition. For all momentum-components where $\sin(p_\mu) \ne 0$
one finds a coupled algebraic equation in $x_\m$.
In all the cases we have studied, it turned out that
the global minimum was either of the form $(p_{\rm min},0,0,0)$
or else of the form $(p_{\rm min},p_{\rm min},p_{\rm min},p_{\rm min})$.)

We first consider an action containing next-nearest neighbors
\beq
 D_{23} = i \sum_\m \g_\m f_3(p_\m) - r\, W_2 + M \,,
\label{d23}
\eeq
\beq
  f_3(p_\m) =  \sin(p_\m)\left[1+c_3(1-\cos(p_\m))\right] \,.
\label{f3}
\eeq
$W_2$ is defined in \seeq{Wn}.  In Table~1 we show the resulting values of
${\rm min}\{2\cosh(\a)\}$ and  ${\rm max}\{\exp(-\a)\}$ for
several values of $c_3$. For each $c_3$ we looked for
the best value of $r$ which we denote $r_{\rm opt}$. One sees that values
of ${\rm max}\{\exp(-\a)\}$  much smaller than 0.5 are feasible.
A plot of $2\cosh(\a)$ for $c_3=4/3$ is shown in Fig.~\ref{var2}.
Notice the flatness of $\cosh(\a)$ for $-1 \le \cos(p_1) \le 0$
at $r=r_{\rm opt}=1.45$.

\begin{table}[bht]
\begin{center}
\vspace*{2mm}
\begin{tabular}{clccc}
\hline
$c_3$ &  $\;\;f_3(p)$ &   $r_{\rm opt}$ &
${\rm min}\{2\cosh(\a)\}$ & ${\rm max}\{\exp(-\a)\}$ \phantom{\Big(} \\
\hline
0     &   $p - {1\over 6} p^3$ &  1.46 &  2.83 &  0.414  \phantom{\Big(} \\
$1/3$ &   $p                 $ &  1.14 &  3.40 &  0.326  \phantom{\Big(} \\
$2/3$ &   $p + {1\over 6} p^3$ &  1.19 &  4.09 &  0.261  \phantom{\Big(} \\
$4/3$ &   $p + {1\over 2} p^3$ &  1.45 &  5.62 &  0.184  \phantom{\Big(} \\
$7/3$ &   $p + p^3           $ &  1.98 &  8.06 &  0.126  \phantom{\Big(} \\
\hline
\end{tabular}
\vspace*{2mm}
\caption{ {\it
${\rm max}\{\exp(-\a)\}$ for the action $D_{23}$ at $M=1$
and for various values of $c_3$ (see text for the definitions).
The second column gives the first two terms
in the expansion of $f_3(p)$. For each $c_3$ we show
the result for $r=r_{\rm opt}$ where ${\rm max}\{\exp(-\a)\}$ is smallest.
}}
\end{center}
\end{table}

We now digress to discuss how the present work relates
to the standard ``improvement program" (see \eg the review~\cite{s}).
In the study of the hadron spectrum,
only a single parameter (the bare quark mass) in the fermion action
needs to be tuned. Once the correct continuum limit has been established,
attention is focused on eliminating those lattice artifacts that
vanish most slowly, that is, linearly with the lattice spacing.
However, in the calculation of weak matrix elements one has
to first establish the correct continuum limit.
This is very problematic with Wilson or staggered fermions
because, due to the loss of full chiral and/or flavor symmetry,
many subtraction coefficients must be tuned.
Controlling those subtractions by having good chiral and flavor
properties simultaneously is thus of higher priority than
the removal of any other lattice error.
Furthermore, in the massless-quark limit
$O(a)$ lattice artifacts are automatically excluded
if chiral symmetry is maintained~\cite{bs}.
In that sense, approaching the chiral limit using domain-wall fermions
encompasses the standard improvement program as well.

\begin{figure}[thb]
\vspace*{0.4cm}
\centerline{
\epsfxsize=8.0cm
\put(-100,0){\epsfbox{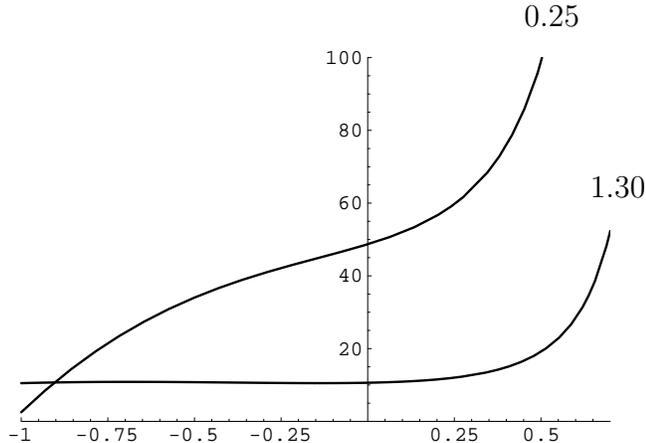}}
\put(95,160){$0.25$}
\put(120,95){$1.30$}
}
\vspace*{0.4cm}
\caption{ \noindent {\it Same as Fig.~\ref{var1} for the action $D_{35}$
with $c_5=2$ for two values of $r$. The optimal value is
$r_{\rm opt}=1.30$, whereas the ``reference value'' is $r_3=0.25$.
Notice the flatness of $\cosh(\a)$ for  $-1 \le x \le 0.25$.
}}
\label{var3}
\vspace*{0.5cm}
\end{figure}

Coming back to the new domain-wall action,
since the new Wilson term $W_2$ starts off at order $p^4$,
the first lattice deviation from a relativistic (tree-level)
dispersion relation comes only from the kinetic term.
This is shown in the second column of Table~1. We observe that while
increasing $c_3$ from zero to $1/3$ improves the dispersion
relation, the opposite is true for $c_3>1/3$.
Although the error is formally of order $a^2$,
it might become significant if $c_3$ is too large.
To gain some idea on the magnitude of the error
consider, say, $p^2 \sim (400\; \mbox{\rm MeV})^2$,
which is relevant for kaon physics, on a lattice with
$a^{-1} \sim 2\; \mbox{\rm GeV}$.
This means $a^2 p^2 \sim 1/25$.
For the last two rows of Table~1, the effect is 2\% and 4\% respectively.

If next-next-nearest neighbors in the same direction are also allowed
one can decrease ${\rm max}\{\exp(-\a)\}$ further
while maintaining a vanishing $p^3$ term. Let
\beq
 D_{35} = i \sum_\m \g_\m f_5(p_\m) - r\, W_3 + M \,,
\label{d35}
\eeq
where again $W_n$ is defined in \seeq{Wn} and where
\beq
  f_5(p_\m) =  \sin(p_\m)
  \left[1 + {1\over 3}(1-\cos(p_\m))+c_5(1-\cos(p_\m))^2\right] \,.
\label{f5}
\eeq
Some values of ${\rm min}\{2\cosh(\a)\}$ and ${\rm max}\{\exp(-\a)\}$
are shown in Table~2. A plot of $2\cosh(\a)$ for $c_5=2$ is
shown in Fig.~\ref{var3}.
Even for the last row in Table~2 ($c_5=50$),
the deviation from Lorentz covariance
is at the level of $(50/4)(a^2 p^2)^2 \sim 2\%$ for
$p^2\sim (400\; \mbox{\rm MeV})^2$.
For $c_5=5$, the deviation is below 2\% up to $(700\; \mbox{\rm MeV})^2$,
and so on.

\begin{table}[bht]
\begin{center}
\vspace*{2mm}
\begin{tabular}{ccccc}
\hline
$c_5$ &  $f_5(p)$ &   $r_{\rm opt}$ &
${\rm min}\{2\cosh(\a)\}$    &   ${\rm max}\{\exp(-\a)\}$ \phantom{\Big(} \\
\hline
1  &   $p +  {13\over 60}\, p^5$ &  0.87 &  \phantom{0}6.85 &  0.149
\phantom{\Big(} \\
2  &   $p +  {14\over 30}\, p^5$ &  1.30 &  10.50 &  0.096 \phantom{\Big(} \\
3  &   $p +  {43\over 60}\, p^5$ &  1.66 &  13.36 &  0.075 \phantom{\Big(} \\
4  &   $p +  {29\over 30}\, p^5$ &  1.96 &  15.74 &  0.064 \phantom{\Big(} \\
5  &   $p + 1{13\over 60}\, p^5$ &  2.21 &  17.74 &  0.057 \phantom{\Big(} \\
10 &   $p + 2{14\over 30}\, p^5$ &  3.18 &  25.44 &  0.039 \phantom{\Big(} \\
30 &   $p + 7{7 \over 15}\, p^5$ &  5.70 &  45.51 &  0.022 \phantom{\Big(} \\
50 &   $p +12{7 \over 15}\, p^5$ &  7.48 &  59.84 &  0.017 \phantom{\Big(} \\
\hline
\end{tabular}
\vspace*{2mm}
\caption{ {\it
${\rm max}\{\exp(-\a)\}$ for the action $D_{35}$
for various values of $c_5$ at $M=1$ and $r=r_{\rm opt}$.
The second column gives the first two terms
in the expansion of $f_5(p)$.
}}
\end{center}
\end{table}

We conclude with a number of technical comments.
Replacing the four-dimensional part of the domain-wall action
by $D_{23}$ ($D_{35}$) approximately doubles (triples)
the number of entries in the fermion matrix.
Therefore one should expect a corresponding increase in the cost of
a single inversion of the fermion matrix at fixed $N_s$.

In the continuum limit, both the standard domain-wall action
and the new actions discussed above support
a single quark (one Weyl field on each boundary) for $|1-M|<1$.
In the case of the standard action there is a four-quark zone
(corresponding to the corner $p_\p$ and its permutations) for
$|3-M| < 1$. When the Wilson term $W_n$ is employed instead,
the four-quark zone is at $|1+r\,2^n-M|<1$.
An additional benefit of the new actions
is that the four-quark and the single-quark zones
are separated by a large gap (as a function of $M$)
for $r \sim r_{\rm opt}$.
We expect that a clear gap should be found in simulations too,
even though its precise location will likely be different
from the weak-coupling limit.

For the standard domain-wall action,
the optimal value of $M$ used in simulations agrees well with
the mean-field estimate of $1+\d M$.
One obtains $\d M$ by substituting a mean value $u$ for each link
variable in the Wilson term.
We will assume that the new actions are gauged in the simplest
way, namely using only products of link variables along straight lines.
(E.g.\ the sites $x$ and $x+2\hat\m$ are connected
via $U_{x,\m}U_{x+\hat\m,\m}$ etc.)
Using a mean link $u \sim 0.8$ at $\b=6.0$ (see \eg \rcite{cppacs}),
the mean-field estimate is $2r(3-4u+u^2) \sim 0.9\, r$ for $D_{23}$ and
$r(10-15u+6u^2-u^3) \sim 1.3\, r$ for $D_{35}$.

Last, for any domain-wall operator with the form of \seeqs{D5} and\seneq{D},
the $\ns\to\infty$ limit defines an overlap-Dirac operator~\cite{ov,nn2}
obeying the Ginsparg-Wilson relation (for a review see \rcite{n}) given by
\beq
  D_{\rm GW} = 1-\g_5\,\e(\g_5 D) \,,
\label{GW}
\eeq
where $\e(x)=\pm 1$ is the sign function acting on each of
the eigenvalues of $\g_5 D$.

\vspace{5ex}
\noindent {\large\bf 4.~~Discussion}
\vspace{3ex}
\secteq{4}

In this paper we showed that the one-loop
wave function of domain-wall fermions behaves like
$s^{-2}\,q_1^s$, where $q_1$ is determined by the free-fermion action.
This is true for a wide class of domain-wall actions,
including those defined in \seeq{D5}.
For the standard action $q_1=0.5$, whereas
the addition of beyond-nearest neighbor couplings allows
for much smaller values of $q_1$.

Only in the weak-coupling limit does $q_1$ fully control the wave function.
For finite $g^2$ up to some critical value $g_c^2$
we expect the wave function to be proportional to some
$\qpt^s$ (up to power corrections) with $\qpt=\qpt(g^2)$.
The $g^2$ dependence can be parametrized in various ways.
In Appendix~B we consider the role of higher-order diagrams,
and the parametrization
\beq
  \qpt^s = q_1^s \exp(s(g^2\,\h_1 + g^4\, \h_2 + \cdots)) \,,
\label{qhigher}
\eeq
is found to be natural.
The Taylor expansion of $\exp(s\,g^2\,\h_1)$
corresponds to a family of 1PI diagrams of all orders,
where the first $\h_1$-dependent terms are two-loop diagrams.
(Analogous statements apply to $\h_2$ etc.)
In terms of $\qpt(g^2)$, one can define $g_c^2$ by the condition
$\qpt(g^2_c)=1$. The existing numerical results suggest
that the (quenched) value of $6/g^2_c$ is very close to 6.0
for the standard domain-wall action.

If both $q_1$ and $\h_1$ were known for a given action,
one could obtain a crude estimate of $g^2_c$ via a linear extrapolation.
We have computed only the $q_1$ values,
so we can only conjecture what trends are likely to affect $g^2_c$.
First, in the lower rows in Table~2, $q_1$ is extremely small.
Nevertheless, if $\h_1$ is large (and positive), $g^2_c$ may end up
being approximately the same as (or, for that matter, even smaller than)
for the standard domain-wall action.  Because of gauge invariance there are
vertices that depend linearly on $c_5$.
Since the lower rows in Table~2 come from actions with numerically
large values of $c_5$ this should, indeed, lead generically
to a large $\h_1$ (and $\h_2$ and so on).

It is therefore safer to focus on the first few rows in Tables~1 and 2,
where one is less prone to the above risk.
The following heuristic argument suggests that, in that range of parameters,
the new domain-wall actions may indeed be superior to the standard one.
When the Wilson parameter is equal to $r_n\equiv 2^{1-n}$,
an action containing the Wilson term $W_n$ (\seeq{Wn})
gives rise to ${\rm min}\{2\cosh(\a)\}=2.5$
(corresponding to $q_1=0.5$) at $p_\p=(\p,0,0,0)$ for $M=1$.
The behavior at $r=r_n$ is therefore a common starting point
over which we may try to improve by increasing $r$. As discussed in Sec.~3,
for the standard action ${\rm min}\{2\cosh(\a)\}$ is relatively
insensitive to $r$. Its largest value (which is 2.8) is obtained
around $r_{\rm opt} \sim 1.3$ -- 1.5, namely $r_{\rm opt}$ is
less than 50\% above $r_1$.
In comparison, for $D_{23}$ at $c_3=4/3$ (Fig.~\ref{var2})
the largest value of ${\rm min}\{2\cosh(\a)\}$ is achieved at
$r_{\rm opt}=1.45$ which is approximately three times $r_2$.
For $D_{35}$ at $c_5=2$ (Fig.~\ref{var3}) the best value
is $r_{\rm opt}=1.3$ which is more than five times $r_3$.

The ability to reach larger values of  ${\rm min}\{2\cosh(\a)\}$
is thus correlated with an enhanced sensitivity to the
Wilson parameter, and with a bigger ratio $r_{\rm opt}/r_n$.
Now, while $\h_1,\h_2,\ldots,$ might in principle grow as $r$ increases from
$r_n$ to $r_{\rm opt}$, it is clear that as
functions of the parameters of the theory their behavior will be
very different from $q_1$ (see Appendix~B).
Therefore it is plausible that there exist ``windows'' of parameters
where $\qpt(g^2)$ is controlled primarily by
the decreasing $q_1$, implying that the exponential suppression
holds up to a larger value of $g^2$.

In Appendix~C.2 we discuss how different ways of approaching the
chiral limit are related to different forms of the spectral-density
function of the (normal-ordered) transfer matrix.
At weak coupling one expects to have a gap, namely almost all eigenvalues
are smaller than some $\l_0 <1$.
The gap region $\l_0 \le \l \le 1$ is not completely devoid of eigenvalues,
but their total number is drastically smaller than just below $\l_0$.
One also expects a big difference between the corresponding
eigenfunctions. Those that lie outside the gap should be continuum-like
modes that spread all over the lattice, while inside the gap
the modes should be highly localized~\cite{scri2,constr}.

As the coupling constant increases a qualitative change takes place.
Near-unity eigenvalues of the transfer matrix proliferate.
For the standard domain-wall action
this is a direct consequence~\cite{nn1,redch} of the proliferation of
approximate zero modes of the hermitian, four-dimensional,
Wilson-Dirac operator~\cite{scri2,scri,constr}.
The change (which seems to take place around
quenched $\b=6.0$) shows the key features
of the phenomenon known in condensed matter as localization~\cite{ml}.
Due to the randomness of generic
gauge-field configurations, in any given part of the lattice there
is a finite probability to find a localized (approximate) zero mode.

Viewing the (hermitian) Wilson-Dirac operator as a hamiltonian,
under its action the fermions can hop only a single site.
But with the new domain-wall actions the relevant
hamiltonian is $\g_5\,D$ (see \seeq{D}).
Now the fermions may hop also two (or three) sites
when the hamiltonian acts just once on a given state.
It should be more difficult to trap the new fermions
inside a small potential well,
as now they have more ways of escaping out of it!
This consideration too suggests that
the critical coupling, where the exponential
suppression is lost, may be larger for the new actions.

The new domain-wall actions considered in this paper
carry with them an obvious extra cost
for a single inversion of the fermion matrix.
One may hope to  reduce chiral symmetry violations also by using
improved gauge actions, because the latter tend to generate
smoother configurations. If this goal is achieved, it may be
a numerically much cheaper way to reduce chiral symmetry violations.
Using the Iwasaki action~\cite{i} it was found that the residual large-$\ns$
pion-mass squared (extrapolated to zero quark mass)
drops by about a factor of two for $a^{-1} \sim 1 \GeV$
\cite{57i}. However, the new residual pion mass is still very big.
Also, in thermodynamics, the Iwasaki
action did not lead to any noticeable reduction
in the residual pion mass (for a detailed discussion
of various improvements see the first paper of \rcite{57i}).

Our analysis suggests a possible explanation why the use
of improved gauge actions has had only a limited success.
If the (tree-level) domain-wall action is
unchanged, the one-loop wave function\seneq{c1} is still controlled by the
same value of $q_1=0.5$. Only the numerical prefactor may change
(cf.\ \seeq{result}). Hence, in this approximation,
the exponential fall-off rate is not getting any better
for improved gauge actions.
(The same reasoning applies to the use of ``fat links''.
One has to be careful, however,
because this argument ignores higher-order corrections, cf.\ \seeq{qhigher}.)

In conclusion, in the one-loop approximation
even a small reduction in $q_1$ leads to a dramatic suppression
of chiral symmetry violations for commonly used values of $\ns$.
If the actual quark's wave function is (even partly)
correlated with the one-loop one, the new actions could give rise
to a significantly better chiral behavior, enough to justify
their increased simulation cost.

\vspace{5ex}
\noindent {\bf Acknowledgements}
\vspace{3ex}

The key results of this paper were worked out during a visit
to Brookhaven National Laboratory last summer.
Extensive discussions with many members of the lattice groups
in Brookhaven Lab.\ and Columbia Univ.\
were essential for the development of this work.
In forming my view on the current state of domain-wall fermion simulations,
as presented in this paper, I benefitted from discussions
with the participants of the Dubna workshop on
``Lattice Fermions and Structure of the Vacuum.''
I thank Tom Blum, Maarten Golterman and Karl Jansen for their useful
comments on this paper, and Karl Jansen also for discussing with me some
as-yet unpublished results. This research is supported in part
by the Israel Science Foundation.

\vspace{5ex}
\centerline{\rule{5cm}{.3mm}}

\newpage
\vspace{5ex}
\noindent {\bf Appendix A. Some technicalities}
\vspace{3ex}
\secteq{A}

In this appendix we collect a few useful formulae.
The inverse tree-level propagator in momentum space is
\beq
  (G^0)^{-1}_{s,t} = i\, \d_{s,t} \sum_{\m} \g_\m \sin(p_\m)
  + W^+_{s,t}(p)\, P_+ + W^-_{s,t}(p)\, P_- \,,
\label{invG0}
\eeq
where
\beq
  W^+_{s,t}(p) = \d_{s+1,t} - B(p) \, \d_{s,t} \,,
\eeq
\beq
  B(p) = 1-M + \sum_\m (1-\cos(p_\m)) \,,
\eeq
and $W^-_{s,t}(p) = W^+_{t,s}(p)$. We consider only the case of
a zero quark mass. Also, for the calculation of the self-energy
we set $M=1$ in the tree-level action (see Sec.~2).
The tree-level propagator was computed in \rcite{sh,pv,at}.
At the corners of the Brillouin zone ($\sin(p_\m)=0$, all $\m$)
the two chiralities decouple in \seeq{invG0}.
For $p \to 0$ the limit is singular. But at the other fifteen corners
the limit is regular, leading to
\beq
  G^0_{s,t} = (W^+)^{-1}_{s,t}(p)\, P_+ + (W^-)^{-1}_{s,t}(p)\, P_- \,.
\label{G0pi}
\eeq
Specifically at $p=p_\p$ and with a semi-infinite coordinate $s,t=0,1,2,\ldots$
one has
\beq
  (W^+)^{-1}_{s,t} = (W^-)^{-1}_{t,s}
  = - {1\over 2} \, \th(t\ge s) \, 2^{s-t} \,,
\eeq
where $\th(t\ge s)=1$ for $t\ge s$ and $\th(t\ge s)=0$ for $t<s$.

The explicit expression for the diagonal self-energy $\S^+_{s,t}$
of \seeq{proj} is given by eq.~(36) of \rcite{at}.
With a slight change of notation it reads
(recall that the external momentum is zero)
\bqry
  \S^+_{s,t} & = &  \int_{-\p}^{+\p} {d^4k\over (2\p)^4}\,
  \Big(4 \sum_\n \sin^2(k_\n/2) \Big)^{-1} \,
  \sum_\m \left(
    \cos^2(k_\m/2) \, (W^+ G^-)_{s,t}
  \phantom{\half} \right.
\NON
  & & \left.
  - \sin^2(k_\m/2) \,  (W^- G^+)_{s,t}
  +  \half \sin^2(k_\m) \,  (G^+ + G^-)_{s,t}
  \right) \,,
\label{mplus}
\eqry
where
\beq
  (G^\pm)^{-1}(p) = \sum_\m \sin^2(p_\m) + W^\pm(p) W^\mp(p) \,.
\eeq
Explicit expressions for $G^\pm_{s,t}$ can be found in refs.~\cite{sh,pv,at}.
As explained in Sec.~2 we are interested in $\S^+_{s,0}$
for $s \gg 1$.
In the saddle-point approximation we set the internal momentum
on the fermion line to $p_\p$ (or its permutations)
in all terms, except in the exponential $\exp(-s'\,\a)$
that occurs inside $G^\pm_{s',0}$,
which is expanded to second order around $p_\p$ using the definition
\beq
  2 \cosh(\a(p)) = {1 + B^2(p) + \sum_\m \sin^2(p_\m) \over B(p)} \,.
\label{alpha}
\eeq
This expansion gives rise to the integrand in \seeq{half}.
Since $\sin(p_\m)=0$ for all $\m$ at $p_\p$,
the last term in \seeq{mplus} is zero.
At $p_\p$ one has $W^\pm\, G^\mp = (W^\mp)^{-1}$.
Since the matrices $W^\pm$ and their inverses are triangular,
the second term in \seeq{mplus} gives zero too.
Only the first term in \seeq{mplus} contributes,
and only for $\m=2,3,4$, leading to \seeq{half}.

We also need the diagonal tree-level propagator $G^{0+}$ (\seeq{proj})
away from the boundaries at $p=0$. For $M\to 1$ and $p \to 0$,
the second-order operators $(W^+ W^-)_{s,t}$ and $(W^- W^+)_{s,t}$
tend to $\d_{s,t}$ (except $(W^- W^+)_{s,t}$ at $t=s=0$).
As a result $G^{0+}_{s,t} = \d_{s-1,t}=W^-_{s,t}$.
(For a semi-infinite $s$-coordinate $W^-$ is a right-inverse of $W^+$.)

Finally we observe that the interacting domain-wall action has a generalized
parity symmetry. Writing the action as
\beq
  \sum_{s,s';\vec{x},\vec{y};x_4,y_4}
  \bar\j_{s,\vec{x},x_4} \,
  D_{s,s';\vec{x},\vec{y};x_4,y_4}(U) \,
  \j_{s',\vec{y},y_4}
\eeq
one has
\beq
  D_{s,s';\vec{x},\vec{y};x_4,y_4}(U)
  = \g_4 \, D_{s',s;-\vec{x},-\vec{y};x_4,y_4}(U') \, \g_4 \,,
\eeq
where
\bqry
  U'_{\vec{x},x_4;4} & = & U_{-\vec{x},x_4;4} \NON
  U'_{\vec{x},x_4;k} & = & U^\dagger_{-\vec{x}-\hat\k,x_4;k}\,,\qquad
  k=1,2,3 \,.
\eqry
This looks like an ordinary parity transformation,
except that we have switched the fifth coordinates of $\j$ and $\bar\j$
($s$ and $s'$). The above discrete symmetry implies
\beq
  \S_{s,t}(\vec{p},p_4) = \g_4 \, \S_{t,s}(-\vec{p},p_4) \, \g_4 \,
\eeq
and when $\sin(p_\m)=0$ for all $\m$, one has
$\tr P_+ \, \S_{s,t} = \tr P_- \, \S_{t,s}$.

\vspace{5ex}
\noindent {\bf Appendix B. Beyond one loop}
\vspace{3ex}
\secteq{B}

In this Appendix we consider the role of higher-order corrections.
Specifically the aim is to show how an $O(g^2)$ correction to $q_1$
is built. The new wave-function is conveniently parametrized as
\bqry
  \d\cpt(s) & \propto & s^{-2}\, \qpt^s \,,
\NON
  \qpt^s = \left( q_1 \exp(g^2 \h_1) \right)^s
  \!\!\!\! & = & \!\!\!
  q_1^s \left( 1 + g^2 \h_1 \, s + \half (g^2 \h_1)^2 \, s^2 + \cdots
  \right) \,.
\label{higher}
\eqry
The last expression suggests that, in the wave function $\d\cpt(s)$,
the $O(g^2)$ correction to $q_1$
arises from a resummation of perturbation theory.
If true, at any finite order
we should find contributions to the wave function whose structure is
$\d\c_1(s)$ times an increasing power of $s$.
(We likewise expect additional $O(g^4)$ \etc corrections to $q_1$,
cf.\ \seeq{qhigher};
the arguments below are, however, too crude to tell
how the power-law part of the wave function depends on $g^2$.)

\setcounter{figure}{0}
\renewcommand{\thefigure}{3\alph{figure}}

\begin{figure}[hbt]
\vspace*{0.8cm}
\centerline{
\epsfxsize=8.0cm
\epsfbox{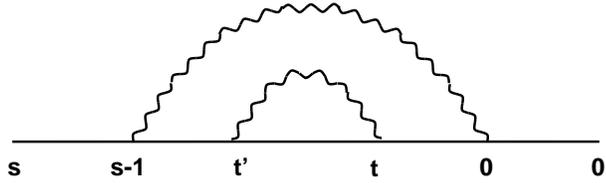}
}
\vspace*{0.0cm}
\caption{ \noindent {\it A two-loop diagram that (likely) contributes
to \seeq{higher}. In this figure and the following ones,
the fifth coordinates correspond to the case
where $1/\sl{p}$ occurs on the rightmost line,
cf.\ \seeqs{G0} and\seneq{c1left}.
}}
\label{two1}
\vspace*{0.5cm}
\end{figure}

We believe that a term $g^2\h\,s\,\d\c_1(s)$ arises from
the two-loop diagrams in Figs.~\ref{two1} and~\ref{two2} (and not from
Figs.~\ref{two3} or~\ref{two4}, see below).
Following Sec.~2, we will consider for definiteness the case
where the rightmost fermion line in Fig.~\ref{two1} corresponds to the singular
part of the tree-level propagator, \seeq{G0}, and therefore
the fifth coordinate of the rightmost vertex is zero.
In Fig.~\ref{two1} the momenta on all three
internal fermion lines can be simultaneously equal (or close) to
$p_\p$ (recall that the external momentum is (close to) zero).
Then, the one-loop exponent
$q_1=\half$ comes with a power $|s-t'|+|t'-t|+t$ which,
for given $s$,
is minimal when the points are ordered: $s \ge t' \ge t \ge 0$.
(Here we have ignored the difference between $s$ and $s-1$
which is negligible for $s \gg 1$.)
When the points are not ordered we obtain an exponentially convergent
series in the excessive length of the fermion's trajectory.
For simplicity we will assume that the points are ordered,
as we are only interested here in arguing that a contribution
proportional to $g^2\,s\,\d\c_1(s)$ exists.
(However, the full series has to be summed in order to obtain the
numerical value of $\h$.)

\begin{figure}[hbt]
\vspace*{0.4cm}
\centerline{
\epsfxsize=8.0cm
\epsfbox{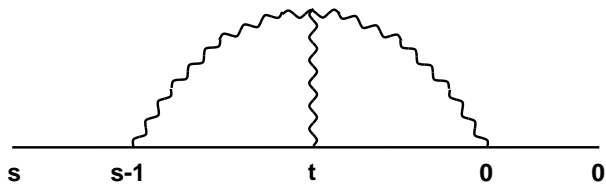}
}
\vspace*{0.4cm}
\caption{ \noindent {\it Another two-loop diagram that contributes
to \seeq{higher}.
}}
\label{two2}
\vspace*{0.5cm}
\end{figure}

Next consider the integration over the loop momentum of
the inner loop. As explained in Sec.~2, this integration should
give rise to a factor of $|t' - t|^{-2}$.
Since the sum $\sum_m m^{-2}$ is convergent,
as a crude approximation one can say that the points $t$ and $t'$
are forced to be close together. Once this extra constraint
has been taken into account, the remaining expression
is independent of $t$ (and $t'$). Therefore the $t$-summation
(approximated by an integral)
gives $\int_0^s dt = s$. The gaussian integration over
the momentum of the outer loop gives rise roughly to the factor $s^{-2}$
(associated with $\d\c_1(s)$) as before. We conclude that
a term proportional to $g^2\,s\, \d\c_1(s)$
may indeed arise from Fig.~\ref{two1}.
(A technical complication is that,
if the two loop-momenta are both equal to $p_\p$,
the momentum flowing through the inner gauge-boson line is zero.
The singularity $(k-p)^{-2}$ is integrable in four dimensions,
but its existence makes the actual calculation quite complicated.
We believe that the above considerations are robust enough to
grasp the dominant behavior of Fig.~\ref{two1}.)

\begin{figure}[hbt]
\vspace*{0.4cm}
\centerline{
\epsfxsize=8.0cm
\epsfbox{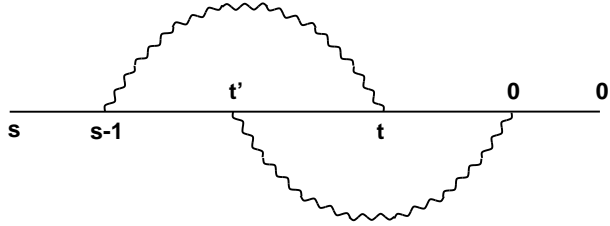}
}
\vspace*{0.4cm}
\caption{ \noindent {\it This two-loop diagram does not contribute
to \seeq{higher}.
}}
\label{two3}
\vspace*{0.5cm}
\end{figure}

\begin{figure}[hbt]
\vspace*{0.4cm}
\centerline{
\epsfxsize=8.0cm
\epsfbox{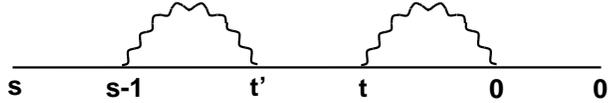}
}
\vspace*{0.4cm}
\caption{ \noindent {\it A reducible two-loop diagram
that does not contribute to \seeq{higher}.
}}
\label{two4}
\vspace*{0.5cm}
\end{figure}

The above argument easily generalizes to the higher-loop diagrams
of Fig.~\ref{three1}. At the $n$th-order one  has $n-1$ ``inner'' loops.
The fifth coordinates are pair-wise close, but otherwise are
constrained only by ordering. Hence one expects a contribution
proportional to
\beq
  \int_0^s dt_1 \int_0^{t_1} dt_2 \ldots \int_0^{t_{n-2}} dt_{n-1}
  = {s^{n-1} \over (n-1)! } \,,
\eeq
in agreement with \seeq{higher}. A similar reasoning implies that
the two-loop and $n$-loop diagrams of Figs.~\ref{two2} and~\ref{three2}
respectively also contribute to the Taylor series in \seeq{higher}.

\setcounter{figure}{0}
\renewcommand{\thefigure}{4\alph{figure}}

\begin{figure}[hbt]
\vspace*{0.4cm}
\centerline{
\epsfxsize=13.0cm
\epsfbox{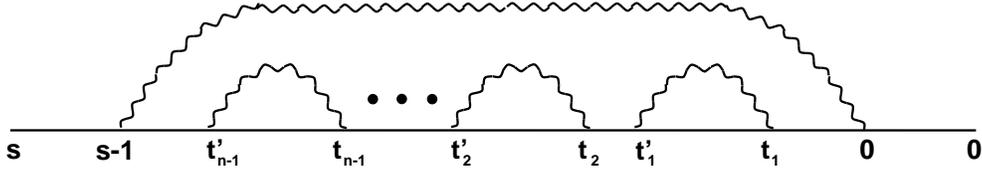}
}
\vspace*{0.4cm}
\caption{ \noindent {\it An $n$-loop diagram that (likely) contributes
to \seeq{higher}.
}}
\label{three1}
\vspace*{0.5cm}
\end{figure}

\begin{figure}[hbt]
\vspace*{0.4cm}
\centerline{
\epsfxsize=13.0cm
\epsfbox{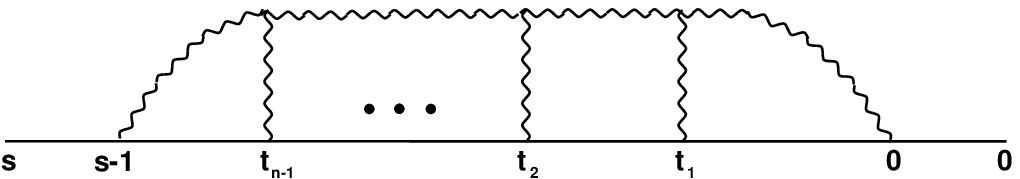}
}
\vspace*{0.4cm}
\caption{ \noindent {\it Another $n$-loop diagram that contributes
to \seeq{higher}.
}}
\label{three2}
\vspace*{0.5cm}
\end{figure}

We now want to explain why the diagrams of Figs.~\ref{two3}
and~\ref{two4} do not contribute to the \rhs of \seeq{higher}.
Consider first the reducible diagram Fig.~\ref{two4}.
The momentum on the middle fermion line is zero
(being equal to the external momentum). Hence the points
$t$ and $t'$ are very close (see Sec.~2 and Appendix~A).
Ignoring the difference between $t$ and $t'$
the two gaussian integrations give rise to the product
$|s-t|^{-2}\, t^{-2}$. The $t$-summation
is then dominated by $t$-values which are either  close to zero or to $s$
(being the fifth coordinates of the rightmost and leftmost vertices).
Therefore the result behaves like $s^{-2}$
(and not like $s^{-2} s = s^{-1}$ as in the case of Fig.~\ref{two1}).

While the diagram in Fig.~\ref{two3} is irreducible,
the momenta on the internal fermion lines cannot be all
equal to $p_\p$. For example, if the momenta
on the first and third internal lines are equal to $p_\p$,
then the momentum on middle line is zero.
Hence, again, the points $t$ and $t'$ must be very close together,
as well as close to either zero or $s$,
and the final result is proportional to $s^{-2}$
as in the case of Fig.~\ref{two4}.

\vspace{5ex}
\noindent {\bf Appendix C. Some non-perturbative observations}
\vspace{3ex}
\secteq{C}

\noindent {\bf C.1.~~Chiral-symmetry restoration}
\vspace{2ex}

We outline here the proof of chiral-symmetry restoration
given in \rcite{fs} in the light of later works
(in particular \rcite{constr,rbbs}). For definiteness we focus on
the anomalous term in the lattice PCAC relation.
(Its vanishing implies that the pion mass will be
zero after taking the infinite-volume and massless-quark limits
in that order.)
The PCAC relation using domain-wall fermions reads~\cite{fs}
\beq
  \D_\m \svev{A^a_{5\m}(x)\, J_5^b(y)}
  = 2\, m_0\, \svev{J^a_{5}(x)\, J_5^b(y)}
  + 2 \, \svev{J^a_{5q}(x)\, J_5^b(y)}
  + \mbox{contact term} \,.
\label{pcac}
\eeq
Here $\D_\m$ is the backward lattice derivative and $m_0$
the bare quark mass.  $A^a_{5\m}(x)$  is the Noether current
of a lattice transformation that assigns opposite charges to fermions
on the half-spaces $0 \le s < \ns/2$ and $\ns/2  \le s < \ns$.
This reduces to a chiral transformation on the quark states,
as long as they are localized in their respective half-spaces.
The pseudo-scalar density $J_5^a(x)$ is composed of fermion variables
situated on the two boundaries, and serves as the standard interpolating
field for pions. The anomalous (lattice-artefact) term in this relation
involves another pseudo-scalar density, $J_{5q}^a(x)$, which is localized
on two $s$-layers exactly half-way in the fifth direction.
(We are assuming an undoubled quark spectrum
(one quark field for each five-dimensional fermion field,
in most applications)
which is true at weak coupling.
As mentioned in the introduction, the massless spectrum
changes at strong coupling~\cite{rbbs} in which case the above
transformations are no longer chiral.)

Before we proceed with the discussion of the anomalous term
we have to address a technical point.
Let $T = \sum_i \sket{i} t_i \sbra{i}$ be the spectral decomposition
of the (positive) first-quantized transfer matrix.
We define a ``normal-ordered'' transfer matrix $Q$
via its spectral decomposition
$Q=\sum_i \sket{i} \l_i \sbra{i}$ where $\l_i={\rm min}\,\{t_i,t_i^{-1}\}$.
Physically, the operation of replacing $t_i >1$ by its inverse
amounts to filling the Dirac sea.
The spectrum of $Q$ lies in the interval $0 \le \l_i \le 1$.
One can show~\cite{redch} that the exact domain-wall propagator
in a given background field is a sum of terms,
each of which involves the matrix $Q$ raised to a positive power
which is a function of the fifth coordinates.

Let us now assume that, for a given background field,
the spectrum of $Q$ lies in the interval $0 \le \l_i \le \l_0$
where $\l_0 < 1$. Coming back to the anomalous term in \seeq{pcac},
for a non-singlet current it involves the propagation of
two fermions over an $s$-separation equal to $\ns/2$ (or $\ns/2 \pm 1$).
The anomalous term is thus bounded by
$(\l_0^{\ns/2})^2=\l_0^\ns$ times a constant.
(Using the second-quantized
transfer matrix formalism one can show that the proportionality
constant is finite, being the norm
of a product of bounded operators~\cite{fs}.)
Moreover, if $Q$ has no eigenvalues larger than
$\l_0$ for all gauge fields
(a condition which is satisfied for a constrained gauge action~\cite{constr})
one finds that the anomalous term falls exponentially
after the functional averaging over the gauge field.

In \rcite{fs} we showed that a very weak bound on the anomalous correlator
exists even if there is no gap at all.
The point is that exact-unity eigenvalues of the transfer matrix
(hence of its normal-ordered version, $Q$, too)
exist only on a submanifold of the lattice gauge-field space
defined by the condition ${\rm det}\, D_{W} = 0$
where $D_W$ is the hermitian four-dimensional Wilson-Dirac operator.
As explained above, when a fermion propagates across an $s$-separation $\ns/2$,
the propagator is bounded by (and, generically, falls like) $\l_0^{\ns/2}$
where $\l_0 \le 1$ is the largest eigenvalue of $Q$.
But $\l_0^{\ns/2}$ is negligible unless $\l_0=1-O(1/\ns)$.
This condition, in turn, will be satisfied only for gauge-field configurations
whose  distance from the above submanifold
does not exceed $O(1/\ns)$.
The volume of the (compact) gauge-field subspace contributing
to the anomalous correlator is therefore finite, and shrinks like $1/\ns$,
implying a similar bound on the correlator itself.
We comment that the restoration of chiral symmetry
is consistent with the fact that the overlap operator
defined by the $\ns \to \infty$ limit (see \seeq{GW}) admits L\"uscher's
chiral symmetry~\cite{ml2} (whose generators are functions
of the gauge field).

\vspace{3ex}
\noindent {\bf C.2.~~Spectral density and effective wave function}
\vspace{2ex}

In the infinite-volume limit one can define a spectral
function $\r_Q(\l)$ associated with  the normal-ordered
transfer matrix $Q$ introduced in the previous subsection,
whose support is (contained in) the interval $[0,1]$.
Here we will not attempt to compute any spectral
function. Instead, we adopt a ``phenomenological'' point of view.
We will assume that a single, continuous, spectral density function $\r(\l)$
has been defined as a suitable configuration average of $\r_Q(\l)$.
Using the considerations of the previous subsection, the aim is to
see how different forms of the spectral function lead
to different ways of approaching the chiral limit.
One can envisage three prototype scenarios which are listed below.
(A recent treatment of domain-wall fermions based on spectral integrals
can be found in \rcite{kik}. A numerical study of a spectral
quantity which is closely related to $\r(1)$ can be found in \rcite{scri}.
See also \rcite{constr}.)

\vspace{1ex}
\noindent 1) {\it Exponential suppression.} Assume that the support of
$\r(\l)$ is the interval $0\le \l \le \l_0$ where $\l_0 < 1$
and that, close to $\l_0$, $\r(\l)$ vanishes
like $(\l_0-\l)^\d$ with $\d>0$.
Consider the propagation of a single fermion from the boundary layer $s'=0$
to some other layer $s$ (We assume $1 \ll s \;\ltap\; \ns/2$).
This involves the integral
\beq
  \int_0^{\l_0} d\l\, \r(\l)\, \l^s
\sim \l_0^s \int_0^{\l_0} d\l\, \r(\l)\,
  \exp\left( s\,{\l-\l_0\over\l_0} \right)
\sim s^{-1-\d}\, \l_0^s \,.
\label{effwf}
\eeq
In order to obtain the power-law correction we have used the assumed
behavior of $\r(\l)$ close to $\l_0$, and wrote
$(\l/\l_0)^s = \exp(s\,\log(1+(\l-\l_0)/\l_0))$.
(The one-loop result of Sec.~2 corresponds to $\l_0 =1/2$ and $\d=1$.)
If $n$ fermions propagate across a similar $s$-interval,
we will obtain the factor $s^{-1-\d} \l_0^s$
for each of them.
We may therefore consider $\ceff(s)=s^{-1-\d} \l_0^s$ as the
effective $s$-coordinate wave function for all quark states.
Since the anomalous divergence $J_{5q}^a(x)$ is a fermion bilinear,
chiral symmetry violations should fall like
$\ceff^2(\ns/2) \sim \ns^{-2(1+\d)} \l_0^\ns$.
(The effective wave function describing the propagation
of $n$ fermions could be somewhat different from $\ceff^n(s)$ due to
interactions between the different particles. Since chiral symmetry violations
are related to $J_{5q}^a(x)$, the relevant effective wave function
is always the one extracted from the sector with one fermion and one
antifermion.)

\vspace{1ex}
\noindent 2) {\it Power-law suppression}. Assume that $\l_0=1$ but
with a vanishing $\r(1)$, namely $\r(\l) \sim (1-\l)^\d$
for $\l \sim 1$ with $\d > 0$.
In that case the result of the spectral integral\seneq{effwf}
will be $s^{-1-\d}$. We might still speak of an effective wave function
$\ceff(s)=s^{-1-\d}$ and expect chiral symmetry violations to fall
like $\ns^{-2(1+\d)}$.

\vspace{1ex}
\noindent 3) {\it (Almost) no suppression}. Last assume that $\l_0=1$
and that $\r(1)$ is non-zero. Remember now the submanifold
discussed in the previous subsection of gauge fields supporting an
eigenvalue one of $Q$. If $\r(1)$ is finite,
configurations close to that submanifold must have a non-negligible
Boltzmann weight. As we have explained, in this case
the only suppression of long-range $s$-correlations  comes
from phase space considerations (the need to pick a configuration located
$O(1/\ns)$ away from that submanifold).
As a result, chiral symmetry violations
fall roughly like $\r(1)/\ns$, and
the concept of a localized, effective wave function breaks down.
($n$-fermion correlations fall like $1/\ns$ too, and not like $1/\ns^n$.)

\vspace{1ex}
For clarity, we have presented above the three mathematically distinct
scenarios. In reality, however, one is likely to encounter a more complicated
behavior, characterized by the existence of a {\it crossover} region.
Let us reexamine the exponential suppression scenario.
As discussed in the previous subsection, in the ensemble of all
gauge-field configurations (as opposed to the case where
the plaquette is constrained to be everywhere small~\cite{constr})
there is a non-zero probability
of finding an eigenvalue arbitrarily close to one, for any $g>0$.
Consider the integrated spectral density $\ci=\int_{\l_0}^1 d\l\,\r(\l)$,
and suppose that $\ci$ is comparable to $\l_0^{N_0}$ for some $N_0$
(up to power corrections).
For  $\ns \;\ltap\; N_0$, chiral symmetry violations will fall
like $\ns^{-2(1+\d)} \l_0^\ns$. The point is that for any $\l_0$
significantly smaller than one, and $N_0$ of the order of (few times) ten,
$\l_0^{N_0}$ will be so small that one will
never have to use $\ns \; \gtap \; N_0$ in a simulation.
For practical purposes this scenario is therefore the same as the
purely-exponential suppression scenario.
If, nevertheless, very large values of $\ns$ will be tried, then
around $\ns \sim N_0$ a crossover to some slower fall-off rate
will be encountered.
(For instance, a crossover to a slower exponential fall-off rate
has been observed in the Schwinger model~\cite{pv}.)

Last consider the relation between the domain-wall
actions discussed in Sec.~3
and the associated overlap operators (see \seeq{GW}).
The spectral function
that controls the approach to the chiral limit of domain-wall fermions
also controls the localization range of the overlap operator.
As is clear from the above discussion, problems start when there is
no gap, namely when
it is not possible to identify a range $\l_0 \le \l \le 1$ (with $\l_0 < 1$)
where the eigenvalue density is (practically) zero.
In that case both the overlap operator and the generators of the
associated L\"uscher symmetries become non-local.
This is the counter-part of the loss of exponential suppression
in the domain-wall case.





\begin{thebibliography}{99}
\bibitem{tom} This possibility was first stressed to me by Tom Blum.
\bibitem{ml} This was pointed to me by Martin L\"uscher.
\end{thebibliography}
\end{document}